\documentclass{emulateapj}

\usepackage{amssymb}

\def\ts     {\thinspace}
\def\kms    {\ts km\ts s$^{-1}$}
\def\etal   {{\rm et\ts al.}}
\def\msol   {$M_{\odot}$}
\def\lsol   {$L_{\odot}$}
\def\bci    {C\ts {\scriptsize I}($^3P_2$$\to$$^3P_1$)}
\def\bcialt    {C\ts {\scriptsize I} $^3P_2$$\to$$^3P_1$}
\def\ci     {C\ts {\scriptsize I}}
\def\cii    {[C\ts {\scriptsize II}]($^3P_{3/2}$$\to$$^3P_{1/2}$)}
\def\ciialt {[C\ts {\scriptsize II}]}
\def\nii    {[N\ts {\scriptsize II}]($^3P_1$$\to$$^3P_0$)}
\def\mgii   {Mg{\scriptsize II}\,($\lambda\lambda$2796,2803\AA )}

\def\aco    {{\rm CO}($J$=1$\to$0)}

\def\cco    {{\rm CO}($J$=3$\to$2)}
\def\ccoalt    {{\rm CO} $J$=3$\to$2}
\def\dco    {{\rm CO}($J$=4$\to$3)}

\def\fco    {{\rm CO}($J$=6$\to$5)}
\def\gco    {{\rm CO}($J$=7$\to$6)}
\def\gcoalt    {{\rm CO} $J$=7$\to$6}

\def\bhcn    {{\rm HCN}($J$=2$\to$1)}

\def\bhco    {{\rm HCO$^+$}($J$=2$\to$1)}

\def\water   {{\rm H$_2$O}($J_{K_{a}K_{c}}$=2$_{12}$$\to$1$_{01}$)}
\def\waterc  {{\rm H$_2$O}($J_{K_{a}K_{c}}$=3$_{13}$$\to$2$_{20}$)}
\def\watera  {{\rm H$_2$O}($J_{K_{a}K_{c}}$=1$_{10}$$\to$1$_{01}$)}
\def\wateralt    {{\rm H$_2$O} $J_{K_{a}K_{c}}$=2$_{12}$$\to$1$_{01}$}

\slugcomment{draft version \today, accepted for publication in the Astrophysical Journal}

\shorttitle{\gco\ and \bci\ Imaging in J1148+5251 ($z$=6.42)}
\shortauthors{Riechers et al.}

\begin{document}

\title{
Imaging Atomic and Highly Excited Molecular Gas in a $z$=6.42 Quasar
Host Galaxy: 
Copious Fuel for an Eddington-Limited Starburst at
the End of Cosmic Reionization}

\author{Dominik A. Riechers\altaffilmark{1,2,7},  
Fabian Walter\altaffilmark{2}, Frank Bertoldi\altaffilmark{3}, Christopher
L. Carilli\altaffilmark{4}, \\ Manuel Aravena\altaffilmark{3}, 
Roberto Neri\altaffilmark{5}, Pierre Cox\altaffilmark{5}, 
Axel Wei\ss\altaffilmark{6}, and Karl M.~Menten\altaffilmark{6}}

\altaffiltext{1}{Astronomy Department, California Institute of
  Technology, MC 249-17, 1200 East California Boulevard, Pasadena, CA
  91125; dr@caltech.edu}

\altaffiltext{2}{Max-Planck-Institut f\"ur Astronomie, K\"onigstuhl 17, 
Heidelberg, D-69117, Germany}

\altaffiltext{3}{Argelander-Institut f\"ur Astronomie, Universit\"at
  Bonn, Auf dem H\"ugel 71, Bonn, D-53121, Germany}

\altaffiltext{4}{National Radio Astronomy Observatory, PO Box O, Socorro, NM 87801}

\altaffiltext{5}{Institut de RadioAstronomie Millim\'etrique, 300 Rue
  de la Piscine, Domaine Universitaire, 38406 Saint Martin d'H\'eres,
  France}

\altaffiltext{6}{Max-Planck-Institut f\"ur Radioastronomie, Auf dem
  H\"ugel 69, Bonn, D-53121, Germany}

\altaffiltext{7}{Hubble Fellow}


\begin{abstract}

We have imaged \gco\ and \bci\ emission in the host galaxy of the
$z$=6.42 quasar SDSS\,J114816.64+525150.3 (hereafter:\ J1148+5251)
through observations with the Plateau de Bure Interferometer.  The
region showing \gco\ emission is spatially resolved, and its size of
5\,kpc is in good agreement with earlier \cco\ observations. In
combination with a revised model of the collisional line excitation in
this source, this indicates that the highly excited molecular gas
traced by the \gcoalt\ line is subthermally excited (showing only
58$\pm$8\% of the \ccoalt\ luminosity), but not more centrally
concentrated.  We also detect \bci\ emission in the host galaxy of
J1148+5251, but the line is too faint to enable a reliable size
measurement.  From the \bci\ line flux, we derive a total atomic
carbon mass of $M_{\rm CI}$=1.1$\times10^{7}$\,M$_\odot$, which
corresponds to $\sim$5$\times$10$^{-4}$ times the total molecular gas
mass.  We also searched for \water\ emission, and obtained a sensitive
line luminosity limit of $L'_{\rm
H_2O}$$<$4.4$\times$10$^9$\,K\,\kms\,pc$^2$, i.e., $<$15\% of the
\cco\ luminosity. The warm, highly excited molecular gas, atomic gas
and dust in this quasar host at the end of cosmic reionization
maintain an intense starburst that reaches surface densities as high
as predicted by (dust opacity) Eddington limited star formation over
kiloparsec scales.

\end{abstract}


\section{Introduction}

Detailed studies of individual quasars at the highest redshifts are
vital to shed light on how today's most massive galaxies formed and to
investigate whether or not the tight correlation between black hole
mass and stellar bulge mass seen today was already in place during the
earliest stages of galaxy formation.  A particularly insightful
example is the very high-redshift quasar J1148+5251 ($z$=6.42), which
is observed at the end of cosmic reionization, only $\sim$870\,Myr
after the Big Bang.\footnote{We use a concordance, flat $\Lambda$CDM
cosmology throughout, with $H_0$=71\,\kms\,Mpc$^{-1}$, $\Omega_{\rm
M}$=0.27, and $\Omega_{\Lambda}$=0.73 (Spergel \etal\
\citeyear{spe07}).} Since its initial discovery in the Sloan Digital
Sky Survey (SDSS; Fan et al.\ \citeyear{fan03}), it has been found to
be an optical point source even at the resolution achieved by the
Hubble Space Telescope (HST; White et al.\ \citeyear{whi05}), with
ultraviolet to infrared colors typical for a type 1 quasar (e.g.,
Jiang et al.\ \citeyear{jia06}). From its bright and broad \mgii\
emission line, a supermassive black hole mass of
3$\times$10$^9$\,\msol\ has been derived (Willott et al.\
\citeyear{wil03}).  It also hosts a 4.2$\times$10$^8$\,\msol\
reservoir of dust that exhibits a high far-infrared (FIR) continuum
luminosity of $L_{\rm FIR}$=2.2$\times$10$^{13}$\,\lsol\ (Bertoldi et
al.\ \citeyear{ber03a}; Beelen et al.\ \citeyear{bee06}).

Subsequently, J1148+5251 was detected in molecular gas emission
(carbon monoxide, CO; Walter et al.\ \citeyear{wal03}; Bertoldi et
al.\ \citeyear{ber03b}), revealing a massive molecular gas reservoir
of $2.4\times$10$^{10}$\,\msol\ that could fuel star formation and
feed the active galactic nucleus (AGN) in this system. High resolution
follow-up observations revealed that this gas is centered on the AGN,
but distributed on scales of 5\,kpc (Walter et al.\
\citeyear{wal04}). This large molecular reservoir harbors a compact 1.5\,kpc 
size region that emits bright emission from the \cii\ interstellar
medium (ISM) cooling line, which is presumably due to active star
formation at an enormous star formation rate (SFR) surface density of
$\sim$1000\,M$_\odot$\,yr$^{-1}$\,kpc$^{-2}$ (Walter et al.\
\citeyear{wal08a}). The dense, star-forming molecular gas component as 
traced by hydrogen cyanide (HCN) is comparatively faint, given the
source's high SFR and FIR luminosity (Riechers et al.\
\citeyear{rie07}).

In addition, it was found that J1148+5251 is radio-quiet and follows
the radio-FIR correlation for star-forming galaxies, providing
additional evidence that the FIR dust emission is dominantly heated by
young stars rather than the AGN (Carilli et al.\ \citeyear{car04};
Wang et al.\ \citeyear{wan08}). To further investigate on which scales
the star formation takes place, we here aimed at resolving the neutral
atomic and the warm, highly excited molecular gas components in the
host galaxy of this unique quasar.

\section{Observations}

\begin{figure}
\epsscale{1.15}
\plotone{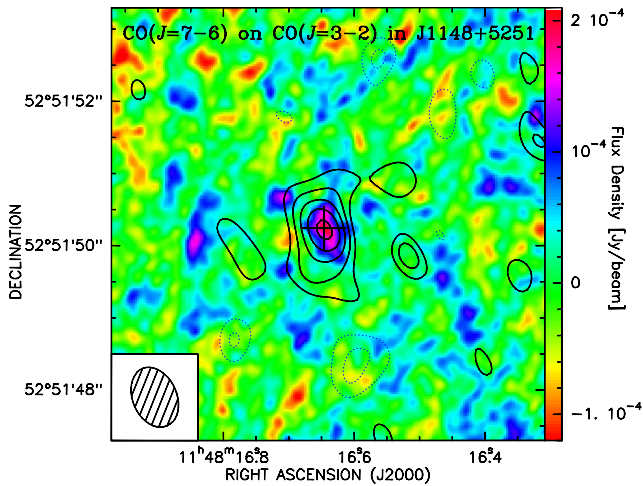}

\caption{Velocity--integrated PdBI map of the \gco\ line emission toward 
J1148+5251 (contours) over the central 414\,\kms, overlayed on the VLA
\cco\ map (Walter et al.\
\citeyear{wal04}; color scale). At a resolution of 0.86$''$$\times$0.61$''$, 
the emission is resolved. The cross indicates the position of the
optical quasar (White et al.\ \citeyear{whi05}). Contours are shown at
(-3, -2, 2, 3, 4, 5, 6)$\times\sigma$ (1$\sigma$ =
0.125\,mJy\,beam$^{-1}$). \label{f1}}
%
\end{figure}

We observed the \gco\ ($\nu_{\rm rest} = 806.6518$\,GHz), \bci\
(809.3435\,GHz), and \water\ (1669.9048\,GHz) emission lines toward
J1148+5251 using the IRAM Plateau de Bure Interferometer (PdBI). At
$z$=6.419, these lines are redshifted to 108.7278, 109.0907, and
225.0849\,GHz (2.8 and 1.3\,mm).  First observations were carried out
with the PdBI's previous generation receivers in the old 6D
configuration in May, June and November 2003 (May observations
published by Bertoldi et al.\
\citeyear{ber03b}), and in April--June and September--November 2004
(23\,tracks total). Further observations were carried out with the new
generation receivers in the new 6A configuration (longest baseline:\
760\,m) in February and March 2007 (7 tracks total). All tracks were
taken under good 3\,mm observing conditions. Of the tracks taken with
dual frequency setup in 2003/2004, 12 were taken under conditions
sufficient for 1\,mm observing. The nearby source 1150+497 (distance
to J1148+5251:\ $3.4^\circ$) was observed every 22.5\,minutes for
pointing, secondary amplitude and phase calibration. For primary flux
calibration, several nearby standard calibrators were observed during
all runs, leading to a calibration that is accurate within 10--15\%
(3\,mm) and 15--20\% (1\,mm), respectively.

Observations with the previous generation receivers were set up using
a total bandwidth of 580\,MHz (single polarization, dual frequency;
corresponding to $\sim$1600\,\kms\ at 2.8\,mm and $\sim$770\,\kms\ at
1.3\,mm). This setup requires two separate tunings at 3\,mm to observe
the \gco\ (14\,tracks) and \bci\ (9\,tracks) emission lines (1\,mm
receivers were tuned to the redshifted \wateralt\ frequency).
Observations with the new generation receivers were set up using a
total bandwidth of 1\,GHz (dual polarization; corresponding to
$\sim$2800\,\kms\ at 2.8\,mm). This setup allows to observe the \gco\
and \bci\ lines simultaneously (selecting a tuning frequency of
108.894\,GHz centered between both lines). All observations were taken
with sufficient bandwidth to cover the underlying continuum
simultaneously.

For data reduction and analysis, the IRAM GILDAS package was used.
All data were mapped using `natural' weighting unless mentioned
otherwise.  The \gco\ data result in a final rms of
0.33\,mJy\,beam$^{-1}$ per 28\,\kms\ channel (resolution:\
1.38$''$$\times$1.17$''$). To optimize the spatial resolution, these
data were also mapped using `uniform' weighting, leading to a
synthesized clean beam size of 0.86$''$$\times$0.61$''$ (4.8\,kpc
$\times$ 3.4\,kpc) and an rms of 0.12\,mJy\,beam$^{-1}$ over
414\,\kms\ (150\,MHz). To maximize the signal--to--noise ratio, the
\bci\ data were tapered to a resolution of 2.78$''$$\times$2.42$''$,
leading to an rms of 0.17\,mJy\,beam$^{-1}$ over 316\,\kms\
(115\,MHz). The 2.8\,mm continuum data (combined from all \gcoalt\ and
\bcialt\ observations) result in a 1.20$''$$\times$0.99$''$ beam
and an rms of 0.046\,mJy\,beam$^{-1}$ over 563.75\,MHz.  The \water\
data result in an rms of 0.76\,mJy\,beam$^{-1}$ over 303\,\kms\
(227.5\,MHz), and the (double sideband) 1.3\,mm continuum data result
in an rms of 0.34\,mJy\,beam$^{-1}$ over 2$\times$563.75\,MHz.

\begin{figure}
\epsscale{1.15}
\plotone{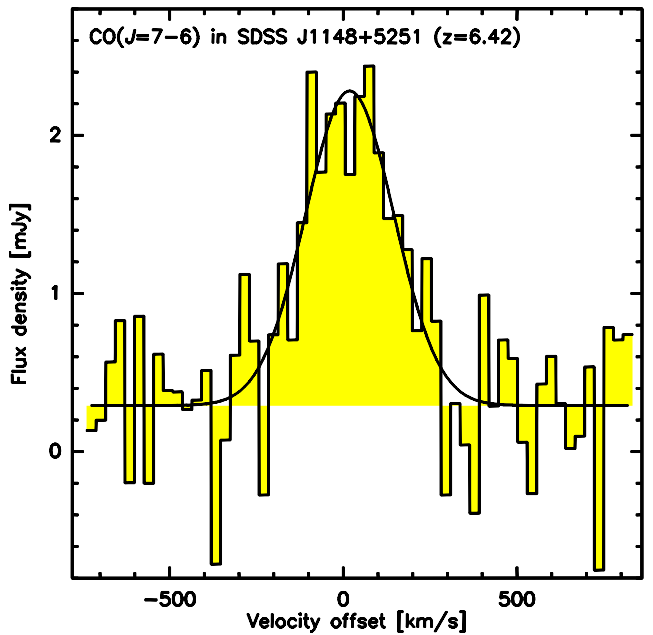}

\caption{PdBI spectrum of the \gco\ and underlying 2.8\,mm continuum 
emission toward J1148+5251 at a resolution of 28\,\kms\ (10\,MHz).
The velocity scale is relative to the redshifted \gco\ frequency at
$z$=6.419. The rms per velocity bin is 0.33\,mJy. The solid line shows
a Gaussian fit to the data. \label{f2}}
%
\end{figure}

\section{Results}

In Figure \ref{f1}, the velocity-integrated \gco\ emission is shown at
a linear resolution of $\sim$4\,kpc (0.7$''$). From elliptical
Gaussian fitting to the 11\,$\sigma$ detection of the source in the
$u-v$ plane, it is found that the emission is spatially resolved on a
scale of 0.9$''$$\pm$0.16$''$ (5.0\,kpc) in the north--south
direction, and marginally resolved (0.54$''$$\pm$0.11$''$; 3.0\,kpc)
in east--west direction. The peak, extent\footnote{Accounting for beam
convolution.} and orientation of the emission are in remarkable
agreement with the lower-excitation \cco\ line emission (color scale
in Fig.~\ref{f1}; Walter et al.\ \citeyear{wal04}), and the peak is
coincident with the position of the optical quasar (cross in
Fig.~\ref{f1}; White et al.\ \citeyear{whi05}). This demonstrates that
the relative astrometry between the \gco\ PdBI and \cco\ VLA
observations is accurate within $<$0.1$''$.  It also indicates that,
even though the \gco\ line is only subthermally excited (see
discussion below), the emission is apparently {\em not} more centrally
concentrated than the \cco\ emission.

In Figure \ref{f2}, the spectrum of the \gco\ emission is shown.  The
line is detected at a peak flux of $S_\nu = 1.99 \pm 0.23$\,mJy and a
width of 297$\pm$35\,\kms, leading to an integrated line flux of
0.63$\pm$0.06\,Jy\,\kms, and a line luminosity of $L'_{\rm
CO(7-6)}$=1.7$\pm$0.2$\times$10$^{10}$\,K\,\kms\,pc$^2$ (see Table
\ref{t1}). The sensitivity of the observations is sufficient to
detect, for the first time, the underlying 2.8\,mm continuum emission
(0.28$\pm$0.07\,mJy). 

\begin{figure}
\epsscale{1.15}
\plotone{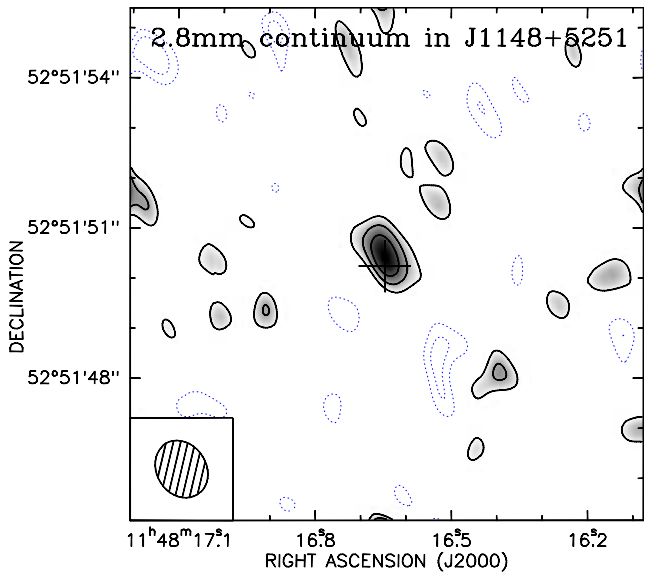}

\caption{Map of the 
2.8\,mm continuum emission toward J1148+5251. At 1.20$''$$\times$0.99$''$ resolution, the emission appears unresolved. Contours are shown at (-3, -2, 2, 3, 4, 5)$\times\sigma$ (1$\sigma$ =
0.046\,mJy\,beam$^{-1}$).
The cross indicates the same position as in Fig.~\ref{f1}. \label{f3}}
%
\end{figure}

Figure \ref{f3} shows a map of the 2.8\,mm continuum emission
integrated over all line-free channels of the {\em combined} CO and
\ci\ observations at $\sim$6\,kpc (1.1$''$) linear resolution. The
continuum emission is detected at a flux level consistent with the
simultaneous line/continuum fit to the \gco\ spectrum within the
errors.

\begin{figure}
\epsscale{1.15}
\plotone{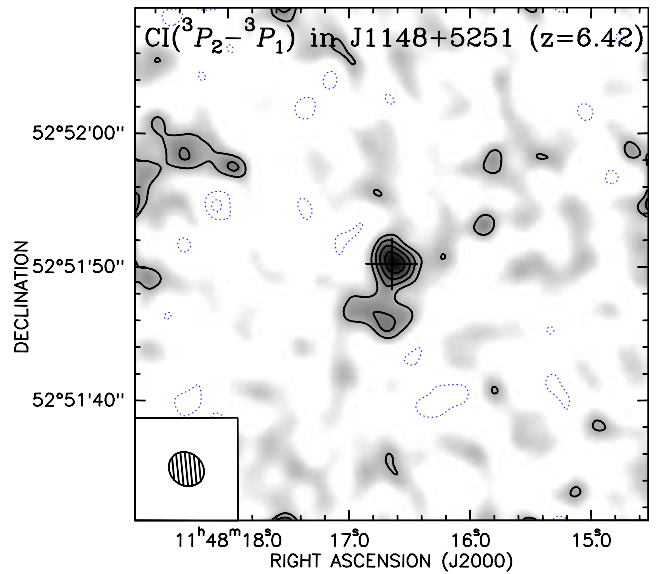}

\caption{Velocity-integrated map of the \bci\ line emission over 316\,\kms\ toward J1148+5251. At
2.78$''$$\times$2.42$''$ resolution, the emission appears unresolved. Contours are shown at (-3, -2, 2, 3, 4, 5)$\times\sigma$ (1$\sigma$ = 0.167\,mJy\,beam$^{-1}$).  The cross indicates the same position as
in Fig.~\ref{f1}. \label{f4}}
%
\end{figure}

In Figure \ref{f4}, a velocity-integrated map of the \bci\ emission
over 316\,\kms\ (i.e., approximately the CO line FWHM) is shown at a
linear resolution of $\sim$15\,kpc (2.6$''$). Emission is detected at
0.98$\pm$0.17\,mJy peak flux, corresponding to a line peak flux of
0.7\,mJy when accounting for the underlying continuum. This
corresponds to an integrated \bci\ line flux of
0.22$\pm$0.05\,Jy\,\kms, and a line luminosity of $L'_{\rm
CI(2-1)}$=6.0$\pm$1.3$\times$10$^9$\,K\,\kms\,pc$^2$. At higher
spatial resolution, the line is still marginally detected, but the
peak flux decreases. This may indicate that the \bci\ emission is
resolved on similar scales as the CO emission, and thus emerges from
the same (molecular) gas phase on global (galactic) scales. However,
this conclusion remains tentative at the present signal--to--noise
ratio.

No evidence for \water\ emission is found in the 1.3\,mm
data. Assuming that the line has the same width as \gco, we derive a
3\,$\sigma$ upper limit of $<$0.69\,Jy\,\kms\ for the integrated line
flux, and a line luminosity limit of \\ $L'_{\rm
H_2O}$$<$4.4$\times$10$^9$\,K\,\kms\,pc$^2$ (i.e., $L'_{\rm
H_2O}$/$L'_{\rm CO(3-2)} < 0.15$).  We clearly detect the underlying
225\,GHz continuum emission at 3.9$\pm$0.8\,mJy.

\section{Discussion}

\subsection{Properties of the Molecular and Atomic ISM in J1148+5251}

\subsubsection{Resolved \gco\ Emission}

We successfully resolved emission from the high--$J$ \gco\ line toward
the $z$=6.42 quasar J1148+5251. The structure and size of the gas
reservoir are consistent with what was derived from previous \cco\
line mapping (Walter et al.\ \citeyear{wal04}).  The line width
derived from our high signal--to--noise \gco\ spectrum is consistent
with that of previous observations of the \fco\ and \gco\ lines
(Bertoldi et al.\ \citeyear{ber03b}). Although the molecular gas
emission appears to be more extended than that from the neutral ISM as
traced by the 158\,$\mu$m \cii\ line, the linewidths and centroids are
consistent to high precision (Maiolino et al.\ \citeyear{mai05};
Walter et al.\ \citeyear{wal08a}).

\subsubsection{Atomic Carbon}

We also detected the upper fine structure line of neutral carbon,
\bci, toward J1148+5251.  This represents the highest $z$ detection of
atomic carbon (or, indeed, any neutral atomic medium) to date, and the
first \ci\ detection in an unlensed high-redshift galaxy.  As the
lower fine structure line (which is redshifted to 66.3\,GHz) cannot be
observed with current facilities, it is not possible to derive an
excitation temperature for \ci. However, for $T_{\rm ex}>20\,$K, the
derived \ci\ mass only weakly depends on $T_{\rm ex}$ (Wei\ss\ et al.\
\citeyear{wei05a}). Assuming an excitation temperature of $T_{\rm
ex}$=36\,K (the average of
the Cloverleaf and IRAS\,F10214+4724, the only high-$z$ galaxies for
which both transitions have been detected; Barvainis et al.\
\citeyear{bar97}; Wei\ss\ et al.\ \citeyear{wei03}, \citeyear{wei05a}; 
Ao et al.\ \citeyear{ao08}), we derive a \ci\ mass of $M_{\rm
CI}$=1.1$\times10^{7}$\,M$_\odot$. From the CO luminosity, a total
molecular gas mass of $M_{\rm H_2}$=2.4$\times10^{10}$\,M$_\odot$ can
be derived (Walter et al.\ \citeyear{wal03}).\footnote{CO in high-$z$
quasar hosts is more widespread than in nearby ULIRGs, but shows
similar physical properties. We thus adopt a low ULIRG CO luminosity
to H$_2$ mass conversion factor of
$\alpha$=0.8\,\msol\,(K\,\kms\,pc$^2$)$^{-1}$ (Downes \& Solomon
\citeyear{ds98}) rather than 
$\alpha$=4--5\,\msol\,(K\,\kms\,pc$^2$)$^{-1}$ as in nearby spirals
(e.g., Scoville \& Sanders \citeyear{sco87a}; Solomon
\& Barrett \citeyear{sol91}). Such low $\alpha$ are also found for $z
\sim 2.5$ submm galaxies (Tacconi et al.\ \citeyear{tac08}).}
We thus find a \ci/H$_2$ mass fraction of 5$\times$10$^{-4}$.
Assuming a [\ci ]/[H$_2$] abundance ratio of 5$\times$10$^{-5}$
(Wei\ss\ et al.\ \citeyear{wei03}, \citeyear{wei05a}), the \ci\ mass
translates to a total molecular gas mass of $M_{\rm
H_2}$=3.7$\times10^{10}$\,M$_\odot$. This is somewhat higher than the
value derived from the CO luminosity, and may indicate either a higher
atomic carbon abundance, or that the CO luminosity to H$_2$ mass
conversion factor is higher than that assumed above [possibly due to
different {\em molecular} abundances relative to nearby ultra-luminous
infrared galaxies (ULIRGs)].

By comparing the line luminosities (in units of \lsol ), one finds
that the \bci\ line provides only $\sim$2\% of the cooling capacity of
the \cii\ line (Maiolino et al.\ \citeyear{mai05}; Walter et al.\
\citeyear{wal08a}); the latter thus clearly remains the strongest
detected coolant of the neutral atomic ISM. This also implies $L_{\rm
CI}/L_{\rm FIR} \simeq 5 \times 10^{-6}$. This ratio is comparable to
those found in nearby starforming and starbursting galaxies, where it
is considered to be a measure for the strength of the non-ionizing
stellar UV radiation field (e.g., G\'erin \& Philips \citeyear{gp00};
Bayet et al.\ \citeyear{bay06}). This supports the assumption that the
bulk of the dust and gas heating in this system is powered by star
formation.  Together with the $L_{\rm [CII]}/L_{\rm FIR}$ ratio of
$\sim 2 \times 10^{-4}$ (Walter et al.\ \citeyear{wal08a}), the above
ratios imply that the UV radiation field in J1148+5251 is relatively
strong, and that the gas density is high [as predicted by
photodissociation region (PDR) models], consistent with nearby ULIRGs
like Arp\,220 (G\'erin \& Philips \citeyear{gp00}, their Fig.~8).

\begin{deluxetable}{ r c c c }
\tabletypesize{\scriptsize}
\tablecaption{Line fluxes and luminosities 
in SDSS\,J1148+5251. \label{t1}}
\tablehead{
& $S_{\nu}{\rm d}v$ & $L'$ & Ref. \\
& [Jy\,\kms ] & [10$^9$\,K\,\kms\,pc$^2$] & }
\startdata
\aco\   & $<$0.11         & $<$143         & 1 \\ 
\cco\   & 0.20 $\pm$ 0.02 & 29.9 $\pm$ 2.7 & 2 \\ 
\fco\   & 0.67 $\pm$ 0.08 & 25.0 $\pm$ 3.0 & 1 \\ 
\gco\   & 0.63 $\pm$ 0.06 & 17.2 $\pm$ 1.8 & 1,3 \\ 
\bhcn\  & $<$0.006        & $<$3.3         & 4 \\
\bhco\  & $<$0.018        & $<$10          & 5 \\
\water\ & $<$0.69         & $<$4.4         & 3 \\
\bci\   & 0.22 $\pm$ 0.05 & 6.0 $\pm$ 1.3  & 3 \\
\cii\   & 3.9 $\pm$ 0.3   & 19.0 $\pm$ 1.6 & 6,7 \\
\nii\   & $<$0.47          & $<$4.0         & 8 \\
\vspace{-2mm}
\enddata 
\tablecomments{${}$ \cco\ corrected by 10\% to account for flux in the outer linewings not covered by the observations. Line fluxes are continuum-corrected where applicable.}
\tablerefs{${}$[1] Bertoldi \etal\ (\citeyear{ber03b}),
[2] Walter \etal\ (\citeyear{wal03}), [3] this work, 
[4] Riechers et al.\ (\citeyear{rie07}), [5] Carilli et al.\ (\citeyear{car05}), 
[6] Maiolino et al.\ (\citeyear{mai05}), [7] Walter et al.\ (\citeyear{wal08a}), 
[8] Walter et al.\ (\citeyear{wal08b}).
}
\end{deluxetable}


\subsubsection{Limits on Water Emission}

We have also searched for \water\ emission toward J1148+5251, but only
obtained a sensitive upper limit. Radiative transfer calculations show
that, assuming the known gas and dust properties of J1148+5251,
\water\ emission is expected to be optically thick over a substantial 
range of the permitted parameter space. Thus, the line luminosity
limit can be directly translated to a surface filling factor limit
(relative to CO $J$=3$\to$2) of $<$15\% (assuming both lines are in
thermal equilibrium).

Sensitive searches for water emission from dense molecular gas were
performed for two other $z$$>$3 galaxies (Riechers et al.\
\citeyear{rie06a}; Wagg et al.\ \citeyear{wag06}), reaching comparable depths.
A search for \waterc\ emission ($\nu_{\rm rest}$=183.310\,GHz) toward
the $z$=3.20 quasar MG\,0751+2716 provided a surface filling factor
limit of $<$6\% relative to \dco\ (assuming the same conditions;
Riechers et al.\ \citeyear{rie06a}). A search for \watera\ emission
($\nu_{\rm rest}$=556.936\,GHz) toward the $z$=3.91 quasar
APM\,08279+5255 provided a surface filling factor limit of $<$12\%
relative to \dco\ (Wagg et al.\ \citeyear{wag06}; Wei\ss\ et al.\
\citeyear{wei07}; Riechers et al.\ \citeyear{rie09}). 

Note that the \waterc\ transition was detected toward the nearby ULIRG
Arp\,220, at about one third of its HCN($J$=1$\to$0) line luminosity
(Cernicharo et al.\ \citeyear{cer06}). This suggests a small surface
filling factor even within dense molecular cloud cores as traced by
HCN. This is, again, assuming optically thick emission in thermal
equilibrium from both lines. Note that the H$_2$O line emission in
Arp\,220 (and the high-$z$ sources) is likely maser-enhanced, which
would even predict substantially smaller surface filling factors
(Cernicharo et al.\ \citeyear{cer06}; Riechers et al.\
\citeyear{rie06a}). Although different H$_2$O and HCN transitions were 
observed toward J1148+5251, the observed \water\ luminosity limit
would be consistent with a H$_2$O/HCN line luminosity ratio similar to
Arp\,220.

\subsection{Continuum Emission}

The new detections of the observed-frame 225\,GHz (rest-frame:
1.67\,THz) and 109\,GHz (807\,GHz) continuum emission are consistent
with the overall dust SED of J1148+5251 (Beelen et al.\
\citeyear{bee06}).  In particular, the detection of 2.8\,mm continuum
emission allows us to better constrain both the \gco\ and \fco\ line
fluxes. From the shape of the SED, we derive an estimated continuum
flux of 0.19\,mJy at the wavelength of the \fco\ emission, by which we
reduce the \fco\ flux relative to the original estimate (Bertoldi et
al.\ \citeyear{ber03b}) in the following.

\section{Modeling and Conclusions}

\begin{figure*}
\epsscale{1.15}
\plotone{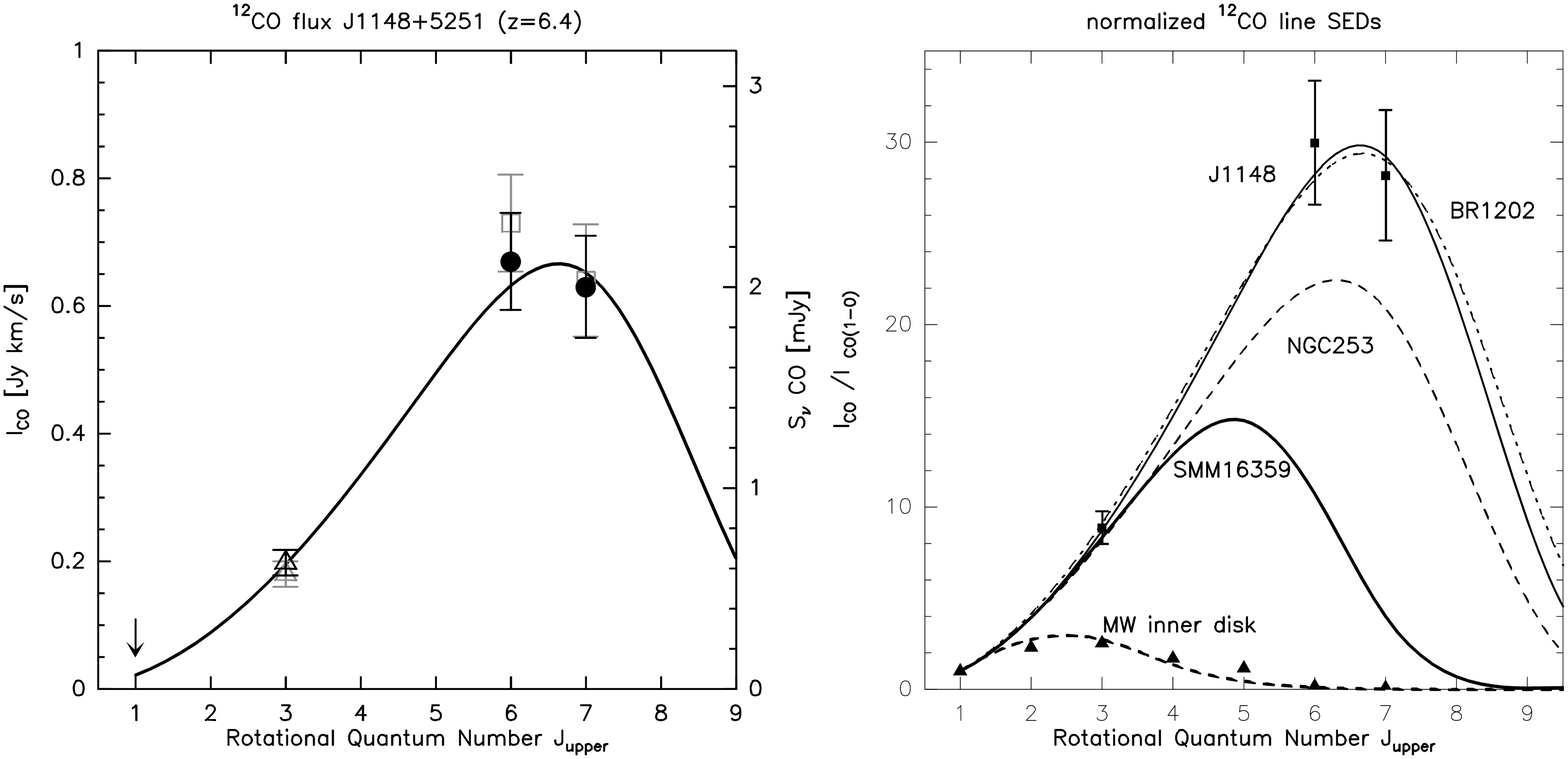}

\caption{{\em Left}:
CO excitation ladder (points) and LVG model (line) for J1148+5251.
The upper limit for \aco\ and the (continuum corrected) data point for
\fco\ are taken from Bertoldi et al.\ (\citeyear{ber03b}). The \cco\ data 
point (corrected for missing flux in the line wings) is taken from
Walter et al.\ (\citeyear{wal03}). The grey points indicate the
original, uncorrected fluxes from Bertoldi et al.\
(\citeyear{ber03b}).  The model predicts kinetic temperatures and gas
densities of $T_{\rm kin} = 50\,$K and $\rho_{\rm gas}({\rm H_2}) =
10^{4.2}\,$cm$^{-3}$. {\em Right}: Comparison of the normalized
excitation ladders for different sources: the inner disk of the Milky
Way (triangles and thick dashed line; Fixsen et al.\
\citeyear{fix99}), the $z$=2.5 submillimeter galaxy SMM\,J16359+6612
(thick solid line; Wei\ss\ et al.\ \citeyear{wei05c}), the nearby
starburst galaxy NGC\,253 (dashed line; G\"usten et al.\
\citeyear{gus06}), the $z$=4.7 quasar BR\,1202-0725 (dash-dotted
line; Riechers et al.\ \citeyear{rie06b}), and J1148+5251 (squares and
solid line).
\label{f5}}
%
\end{figure*}

\subsection{CO Line Excitation}

Given the high signal--to--noise of the \gco\ observations and the
fact that we successfully detected the (observed-frame) 2.8\,mm
continuum emission (which also provided an updated \fco\ line flux),
we have calculated new Large Velocity Gradient (LVG) models of the
line excitation in this system (see Bertoldi et al.\ \citeyear{ber03b}
for the original study). In this LVG model study, the kinetic gas
temperature and density are treated as free parameters.  For all
calculations, the H$_2$ ortho--to--para ratio was fixed to
3:1,\footnote{This is due to the relative statistical weights of the
symmetrical (ortho) and antisymmetrical (para) eigenstates of the
wavefunction: there are three symmetrical combinations of the spins of
both H nuclei, but there is only one antisymmetrical combination.}
the cosmic microwave background temperature was fixed to 20.25\,K (at
$z$=6.42), and the Flower (\citeyear{flo01}) CO collision rates were
used. We adopted a CO abundance per velocity gradient\footnote{Only
the ratio of [CO] and ${\rm d}v/{\rm d}r$ enters the LVG
calculations. Thus, our solutions can account for the factor of a few
variations in [CO] found between nearby star-forming galaxies by
adjusting the `internal' parameter ${\rm d}v/{\rm d}r$ accordingly.
Here, [CO]/(${\rm d}v/{\rm d}r$) is fixed to that of the nearby
starburst M82 (Wei\ss\ et al.\ \citeyear{wei05b}).} of [CO]/(${\rm
d}v/{\rm d}r) = 1 \times 10^{-5}\,{\rm pc}\,$(\kms)$^{-1}$ (e.g.,
Wei\ss\ et al.\ \citeyear{wei05b}, \citeyear{wei07}; Riechers et al.\
\citeyear{rie06b}), and a CO disk radius of 2.5\,kpc as found from the
spatially resolved \cco\ and \gco\ observations. The best solution was
obtained for a spherical, single-component model with a CO disk
filling factor of 0.16, $T_{\rm kin}$=50\,K, and $\rho_{\rm
gas}$(H$_2$)=10$^{4.2}$\,cm$^{-3}$ (Fig.~\ref{f5}). These values are
comparable to those found for other high-$z$ quasars (e.g., Riechers
et al.\ \citeyear{rie06b}; Wei\ss\ et al.\ \citeyear{wei07b}). This
model predicts that the \cco\ line emission is thermalized.  Also, the
model-predicted gas temperature is consistent with that of the dust
(55 $\pm$ 5\,K; Beelen et al.\ \citeyear{bee06}).

\subsection{Star Formation Rate and its Surface Density}

The dust reservoir in J1148+5251 exhibits a FIR continuum luminosity
of $L_{\rm FIR}$=2.2$\times$10$^{13}$\,\lsol\ (Bertoldi et al.\
\citeyear{ber03a}; Beelen et al.\ \citeyear{bee06}), about half of which
emerges from a compact, 0.75\,kpc radius region (Walter et al.\
\citeyear{wal08a}). Its molecular gas, dense gas, radio continuum an dust
properties are consistent with $L_{\rm FIR}$ being dominantly powered
by star formation (e.g., Beelen et al.\ \citeyear{bee06}; Riechers et
al.\ \citeyear{rie07}; Walter et al.\ \citeyear{wal08a}). Assuming the
AGN contribution to $L_{\rm FIR}$ is small, we thus
derive\footnote{Assuming SFR=1.5$\times$10$^{-10}\,L_{\rm
FIR}$(\msol\,yr$^{-1}$/\lsol) (Kennicutt \citeyear{ken98}).} an
integrated SFR of 3300\,\msol\,yr$^{-1}$, about half of which takes
place within a 0.75\,kpc radius region.  This high SFR is consistent
with the high luminosity of the \cii\ ISM cooling line (Maiolino et
al.\ \citeyear{mai05}), which emerges from a region of 0.75\,kpc
radius within the molecular gas reservoir (Walter et al.\
\citeyear{wal08a}). An intense, kpc-scale starburst is also needed to
explain the fact that the molecular gas is warm and highly excited
over a large, 5\,kpc size region (as shown by the new, spatially
resolved CO $J$=7$\to$6 data), which requires a heating source of
similar size (given that radiation from a compact heating source such
as the AGN likely cannot penetrate the dense molecular reservoir out
to such large scales).

Thus, the (peak) size and brightness of the resolved \cii\ line and
FIR continuum emission are consistent with a nuclear `hyper'-starburst
as the main heating source for the gas and dust, i.e., a starburst
that exhibits a flux of $F_{\rm FIR}$=10$^{13}$\,\lsol\,kpc$^{-2}$,
produced by an enormous SFR surface density of $\Sigma_{\rm
SFR}=1000$\,\msol\,yr$^{-1}$\,kpc$^{-2}$ (Walter et al.\
\citeyear{wal08a}). Such high $F_{\rm FIR}$ and $\Sigma_{\rm
SFR}$ are typically observed in the most extreme star-forming
environments in the local universe. As a reference, the Galactic star
forming cloud Orion exhibits a FIR luminosity of
1.2$\times$10$^5$\,\lsol\ within its central arcmin$^2$
(0.013\,pc$^2$; Werner et al.\ \citeyear{wer76}). This corresponds to
a $F_{\rm FIR}$ of $\sim$10$^{13}$\,\lsol\,kpc$^{-2}$. Nearby ULIRGs
exhibit a FIR luminosity of typically 3$\times$10$^{11}$\,\lsol\
within their most active $\sim$100\,pc size regions (Downes \& Solomon
\citeyear{ds98}). This, again, corresponds to
a $F_{\rm FIR}$ of $\sim$10$^{13}$\,\lsol\,kpc$^{-2}$. $F_{\rm FIR}$
(or $\Sigma_{\rm SFR}$) as extreme as in J1148+5251 thus are observed
in the Galaxy and the local universe, albeit in areas that are by 2--8
orders of magnitude smaller. Over surface areas comparable to
J1148+5251, extreme starbursts at $z>2$ (submillimeter galaxies; SMGs)
have typical $\Sigma_{\rm SFR}$ of
$\sim$80\,\msol\,yr$^{-1}$\,kpc$^{-2}$ (Tacconi et al.\
\citeyear{tac06}), which are by about an order of magnitude smaller.

Intriguingly, recent high resolution molecular gas and radio
continuum observations of the $z$=4.4 merging quasar host of
BRI\,1335-0417 indicate $\Sigma_{\rm SFR}$ comparable to those in
J1148+5251 (assuming that the FIR-emitting dust is not more extended
than the molecular gas; Riechers et al.\ \citeyear{rie08a}; Momjian et
al.\ \citeyear{mom07}).

\subsection{Radiation-Pressure Supported Starburst Disks}

A SFR surface density of $\Sigma_{\rm
SFR}=1000$\,\msol\,yr$^{-1}$\,kpc$^{-2}$ is consistent with theories
of `maximum starbursts' (Elmegreen \citeyear{elm99}). At such high
surface densities, feedback from star formation is likely to occur,
which may (self-)regulate the starburst. If the maximum intensity of a
radiation pressure-supported, galaxy scale starburst is determined by
the Eddington limit for dust, a number of characteric limits arise in
these `maximum starburst' theories (Thompson et al.\
\citeyear{tho05}). Assuming a Salpeter-like initial mass function
(IMF), a constant gas-to-dust ratio with radius, and that the disk is
self-regulated (i.e., Toomre-$Q$$\sim$1), such an Eddington-limited
starburst has $\Sigma_{\rm SFR} \sim
1000$\,\msol\,yr$^{-1}$\,kpc$^{-2}$, $F_{\rm FIR} \sim
10^{13}$\,\lsol\,kpc$^{-2}$, and an effective temperature of 88\,K
(using equations 34--36 of Thompson et al.\ \citeyear{tho05}). As
shown above, J1148+5251 has $\Sigma_{\rm SFR}$ and $F_{\rm FIR}$ close
to these theoretical limits over a 0.75\,kpc radius region.

For a flat disk as assumed here, $F_{\rm FIR}$=$L_{\rm
FIR}$/($\pi\,r^2)$=$\sigma_{\rm SB}\,T_{\rm eff}^4$, where $r$ is the
disk radius, and $\sigma_{\rm SB}$ is the Stefan-Boltzmann
constant.\footnote{Assuming a black-body, which the dust SED only
approximates.} About half of $L_{\rm FIR}$ comes from a 0.75\,kpc
radius region in J1148+5251. This corresponds to $T_{\rm eff}^{\rm
peak}$=82\,K, which is also consistent with the limit. Note that,
integrated over the whole 5\,kpc region traced by CO (and full $L_{\rm
FIR}$), this simplified calculation predicts $T_{\rm eff}$=53\,K,
which is in good agreement with the kinetic temperature of the
molecular gas as derived above, and the dust temperature obtained from
fitting the SED of the source.

The vertical optical depth of a disk is given by
$\tau_{V}$=$\Sigma_{\rm gas}\,\kappa/2$, where $\Sigma_{\rm gas}$ is
the gas surface density, and $\kappa \simeq \kappa_0\,T_{\rm eff}^2$
(valid for $T_{\rm eff} \lesssim 200$\,K; $\kappa_0 \simeq 2.4 \times
10^{-4}$\,cm$^2$\,g$^{-1}\,$K$^{-2}$; see Thompson et al.\
\citeyear{tho05}) is the Rosseland mean opacity for dust.
The assumption of a constant gas-to-dust ratio with radius implies
that about half of the gas mass is found within a 0.75\,kpc radius
region in J1148+5251. Together with the $T_{\rm eff}^{\rm peak}$
derived above, this implies $\tau_{V}$=1.03 (i.e., moderately
optically thick), consistent with the assumptions of the
Eddington-limited starburst model.

From the dynamical \cii\ line map of J1148+5251, we can derive a
dynamical mass of $M_{\rm dyn}$$\sim$1.5$\times$ 10$^{10}$\,\msol\
within a 0.75\,kpc radius region. Assuming that $M_{\rm dyn}$ traces
the total mass within that region, this corresponds to a gas mass
fraction of $f_{\rm gas}$=0.72. Using equation 37 of Thompson et al.\
(\citeyear{tho05}), this suggests that the disk is optically thick out
to a radius of $\sim$400\,pc.\footnote{As $M_{\rm
dyn}$$\propto$$\sigma_{\rm FWHM}^2$ was used to determine the gas
fraction, this value is independent of the assumed disk inclination.}
This value is by almost a factor of 2 smaller than 750\,pc. This may
indicate that the potential is not isothermal, as assumed in the
model. It may also indicate that the three-dimensional distribution of
the gas and dust is more complicated than assumed here. However, the
model reproduces a number of the observed properties of J1148+5251
fairly well. Our observations thus are, indeed, consistent with a
kiloparsec-scale `maximum starburst', radiating close to its
characteristic Eddington limit. This hyper-starburst is harbored by a
large, 5\,kpc size, overall warm and highly excited molecular gas
reservoir that hosts $2.4 \times 10^{10}$\,\msol\ of molecular
hydrogen and 1.1$\times$10$^7$\,\msol\ of atomic carbon.

\subsection{Dense Molecular Gas and Gas Surface Density}

\begin{figure}
\epsscale{1.2}
\plotone{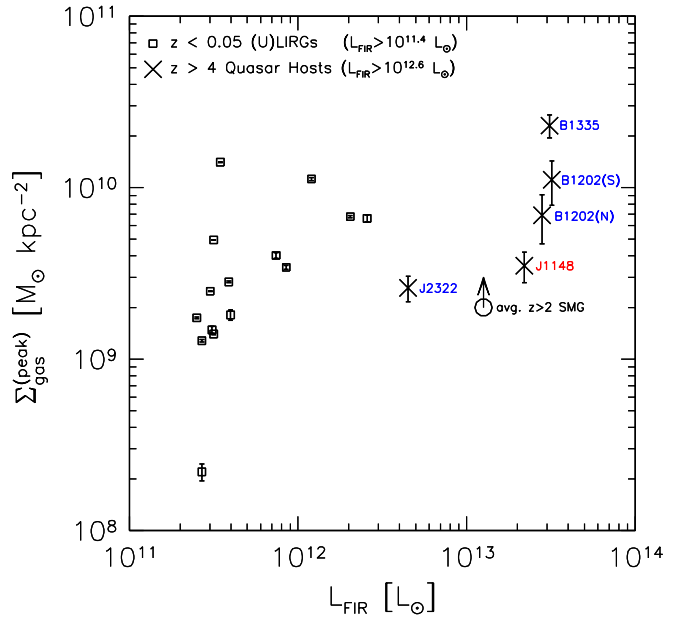}

\caption{Molecular gas peak surface density ($\Sigma_{\rm gas}^{\rm (peak)}$) 
as a function of FIR luminosity ($L_{\rm FIR}$) for nearby luminous
and ultra-luminous infrared galaxies (squares; Wilson et al.\
\citeyear{wil08}) and $z>4$ quasar host galaxies (crosses), including 
PSS\,J2322+1944 ($z$=4.12), BRI\,1335-0417 ($z$=4.41), BR\,1202-0725
north and south ($z$=4.69), and J1148+5251 ($z$=6.42; this work;
Carilli et al.\ \citeyear{car02}; Riechers et al.\ \citeyear{rie08a},
\citeyear{rie08b}).  Data for J2322+1944 are corrected for
gravitational lensing. For comparison, the surface-averaged
$\Sigma_{\rm gas}^{R_{1/2}}$ at half intensity radius for $z>2$ SMGs
is shown (circle; sample average of Tacconi et al.\ \citeyear{tac06}).
\label{f6}}
%
\end{figure}

There are additional lines of evidence that suggest that $\Sigma_{\rm
SFR}$ as high as 1000\,\msol\,yr$^{-1}$\,kpc$^{-2}$ on kpc scales (as
observed in J1148+5251) are not unreasonable. Recent investigations of
the (in the nearby universe linear) HCN-FIR luminosity correlation at
high $z$ (including J1148+5251) show tentative evidence for an excess
in $L_{\rm FIR}$ relative to the observed $L'_{\rm HCN}$ toward the
highest $L_{\rm FIR}$ systems (both with and without AGN; Carilli et
al.\ \citeyear{car05}; Gao et al.\ \citeyear{gao07}; Riechers et al.\
\citeyear{rie07}). Such an excess can be explained by higher median
gas densities (comparable to the critical density of HCN $J$=1$\to$0)
in the most FIR-luminous galaxies (Krumholz \& Thompson
\citeyear{kt07}; Narayanan et al.\ \citeyear{nar08a}). 
Even when assuming a constant efficiency of turning a unit mass of
molecular gas into stars among all galaxies, this would imply a higher
SFR per unit area for the most FIR-luminous systems.

The (molecular) gas surface density in J1148+5251 consistently reaches
high peak values of up to $\Sigma_{\rm gas}^{\rm (peak)} = 3.5 \times
10^9$\,\msol\,kpc$^{-2}$. In Figure~\ref{f6}, the $\Sigma_{\rm
gas}^{\rm (peak)}$ of J1148+5251, other $z>4$ quasar host galaxies
(Carilli et al.\ \citeyear{car02}; Riechers et al.\ \citeyear{rie08a},
\citeyear{rie08b}), and a sample of nearby LIRGs and ULIRGs (Wilson et
al.\ \citeyear{wil08}) is shown as a function of $L_{\rm FIR}$. For
comparison, a lower limit for $z>2$ SMGs is shown (Tacconi et al.\
\citeyear{tac06}). Wilson et al.\ (\citeyear{wil08}) suggested that
there may be a trend of increasing $\Sigma_{\rm gas}^{\rm (peak)}$
with $L_{\rm FIR}$. This would be consistent with a higher density of
gas clouds toward more intensely star-forming galaxies. However, their
sample of nearby galaxies spans only about an order of magnitude in
$L_{\rm FIR}$. Including the more FIR-luminous high-$z$ systems, this
trend gets substantially weaker. In fact, the high-$z$ systems appear
to occupy a similar range in $\Sigma_{\rm gas}^{\rm (peak)}$ as the
nearby ULIRGs, and the relation appears to saturate close to the
typical gas surface densities of Giant molecular clouds (GMCs). This
would imply that, in the most FIR-luminous systems, the entire,
kiloparsec-scale regions with the highest $\Sigma_{\rm gas}^{\rm
(peak)}$ have gas surface densities close to those in GMCs. One
limitation of this conclusion is that the nearby galaxies were studied
at a factor 1.8 higher linear (physical) resolution than the $z>4$
systems. This, however, is not sufficient to explain the weakening of
the trend. In particular, the $z$=4.4 quasar BRI\,1335-0417 with the
highest $\Sigma_{\rm gas}^{\rm (peak)}$ (Riechers et al.\
\citeyear{rie08a}) was studied at only 20\% lower resolution than the
nearby sample average. Even assuming conservatively that $\Sigma_{\rm
gas}^{\rm (peak)}$ is by $\sim$50\% higher in this source (according
to the difference in beam area), it is still consistent with that in
the nuclei of the nearby ULIRG Arp\,220 (Downes \& Solomon
\citeyear{ds98}). This would be consistent with a picture in which the
molecular ISM in ULIRGs and even more FIR-luminous systems is
dominantly in a rather dense, relatively continuous phase (rather than
in individual, moderately dense molecular clouds), exhibiting
optically thick emission. While the measured (peak) gas surface
densities are comparable, the gas is spread out over larger areas in
the high-$z$ quasar host galaxies relative to the (dense) nuclei of
nearby ULIRGs, consistent with the higher total gas masses and SFRs,
and higher median gas densities (integrated over the whole
galaxies). Overall, this may suggest that a tighter correlation would
be found if $\Sigma_{\rm gas}^{\rm (peak)}$ is compared to
$\Sigma_{\rm SFR}$ instead of the integrated $L_{\rm FIR}$ (as a proxy
of SFR).

\subsection{Evolution of the Gaseous and Stellar Components}

Assuming that the massive amounts of molecular material available on
5\,kpc scales in J1148+5251 are converted into stars at an efficiency
of 5\%--10\% as in GMC cores (e.g., Myers et al.\ \citeyear{mye86};
Scoville et al.\ \citeyear{sco87}), the SFR corresponds to a gas
depletion timescale of only (0.7--1.4) $\times$ 10$^8$\,yr. This is
about one-third of the value found for BRI\,1335-0417 (Riechers et
al.\ \citeyear{rie08a}), suggesting a rapid buildup of the stellar
component in both galaxies. Still, this gas depletion timescale is a
few times larger than the dynamical and free-fall times of both the
0.75\,kpc radius region traced by \ciialt, and the full 5\,kpc size
molecular reservoir. In this regard, the ongoing star formation may
still be considered slow.

The \cii\ dynamical mass estimate suggests that J1148+5251 hosts a
total mass of $\sim$1.5 $\times$ 10$^{10}$\,\msol\ within a
$\sim$0.75\,kpc radius region.
A present day, massive elliptical galaxy with a black hole mass
comparable to J1148+5251 and a velocity dispersion of 300\,\kms\ hosts
a total mass of $\sim$5--8 $\times$ 10$^{10}$\,\msol\ within a central
region of the same size (and a total stellar mass of $\sim$2 $\times$
10$^{12}$\,\msol\ distributed over its entire spheroid; e.g., H\"aring
\& Rix \citeyear{hr04}). Even assuming that the whole amount of
molecular material in J1148+5251 were to eventually contribute to the
assembly of the stellar spheroid within this small central region
leaves a factor of a few difference between the central mass budgets
of the $z$=6.42 quasar and a $z$=0 massive elliptical galaxy. This
suggests that J1148+5251 has to accrete additional material by $z$=0
(e.g., through subsequent mergers) to grow a spheroid mass comparable
to its (likely) present-day counterparts.

On the other hand, the widths of the \gco\ and \cii\ lines are the
same within the uncertainties, while the sizes of the CO and
\ciialt\-emitting regions appear to be different by a factor of
$\sim$3. Thus, [CII] traces a smaller dynamical mass than CO,
corresponding to a shallower potential well. This difference may
indicate that the gas in J1148+5251 has not yet fully coalesced,
possibly due to an ongoing merger. This would be consistent with the
high star formation efficiency found in the central region of this
galaxy (Walter et al.\ \citeyear{wal08a}), and the comparatively high
fraction of dense gas (Riechers et al.\ \citeyear{rie07}).

We conclude that the highly excited, several kpc scale size gas
reservoirs in dust- and gas-rich high-$z$ quasar host galaxies are
dominantly heated by the large-scale starbursts that they maintain,
consistent with cosmological simulations of merger-driven $z \sim 6$
quasar formation (Li et al.\ \citeyear{li08}; Narayanan et al.\
\citeyear{nar08b}). These starbursts reach surface densities as
predicted by Eddington-limited star formation over kpc scales,
accompanied (and probably supported) by ongoing major, `wet' merger
activity in some cases.

\acknowledgments 
We thank Todd Thompson, Arjen van der Wel, and Hans-Walter Rix for
helpful discussions, the anonymous referee for helpful comments, and
Christian Henkel for his LVG code. This research is based on
observations carried out with the IRAM PdBI.
IRAM is supported by INSU/CNRS (France), MPG (Germany), and IGN
(Spain). DR acknowledges support from from NASA through Hubble
Fellowship grant HST-HF-01212.01-A awarded by the Space Telescope
Science Institute, which is operated by the Association of
Universities for Research in Astronomy, Inc., for NASA, under contract
NAS 5-26555.  CC acknowledges support from the Max-Planck-Gesellschaft
and the Alexander von Humboldt-Stiftung through the
Max-Planck-Forschungspreis 2005. DR \& FW appreciate the hospitality
at the Aspen Center for Physics, where part of this manuscript was
written.

\end{document}